\documentclass[english,
 journal=ancac3,
 manuscript=article,
]{achemso}

\usepackage[version=3]{mhchem} 

\usepackage{xr}
\makeatletter
\newcommand*{\addFileDependency}[1]{
   \typeout{(#1)}
   \@addtofilelist{#1}
   \IfFileExists{#1}{}{\typeout{No file #1.}}
}
\makeatother

\newcommand*{\myexternaldocument}[1]{
     \externaldocument{#1}
     \addFileDependency{#1.tex}
     \addFileDependency{#1.aux}
}
\myexternaldocument{SI}

\usepackage{chemmacros}
\usepackage{upgreek}
\usepackage{amstext}
\usepackage{graphicx}
\usepackage{gensymb}
\usepackage{babel}

\let\oldAA\AA
\renewcommand{\AA}{\text{\normalfont\oldAA}}

\usepackage[figurename=]{caption}

\usepackage[section]{placeins}

\makeatletter
\let\ftype@table\ftype@figure
\makeatother

\makeatletter

\title{Room temperature single-photon superfluorescence from a single epitaxial cuboid nano-heterostructure}

\author{John P. Philbin}
\affiliation{Harvard John A. Paulson School of Engineering and Applied Sciences, Harvard University, Cambridge, MA 02138, USA}
\email{jphilbin@g.harvard.edu}

\author{Joseph Kelly}
\affiliation{Department of Chemistry, Stanford University, Stanford, CA 94305, USA}

\author{Lintao Peng}
\affiliation{Center for Nanoscale Materials, Argonne National Laboratory, Lemont, IL 60439, USA}

\author{Igor Coropceanu}
\affiliation{Department of Chemistry and James Franck Institute, University of Chicago, IL 60637, USA}

\author{Abhijit Hazarika}
\affiliation{Department of Chemistry and James Franck Institute, University of Chicago, IL 60637, USA}

\author{Dmitri V. Talapin}
\affiliation{Center for Nanoscale Materials, Argonne National Laboratory, Lemont, IL 60439, USA}
\alsoaffiliation{Department of Chemistry and James Franck Institute, University of Chicago, IL 60637, USA}

\author{Eran Rabani}
\affiliation{Department of Chemistry, University of California, Berkeley, CA 94720, USA}
\altaffiliation{Materials Sciences Division, Lawrence Berkeley National Laboratory, Berkeley, CA 94720, USA}
\alsoaffiliation{The Sackler Center for Computational Molecular and Materials Science, Tel Aviv University, Tel Aviv 69978, Israel}

\author{Xuedan Ma}
\affiliation{Center for Nanoscale Materials, Argonne National Laboratory, Lemont, IL 60439, USA}
\email{xuedan.ma@anl.gov}

\author{Prineha Narang}
\affiliation{Harvard John A. Paulson School of Engineering and Applied Sciences, Harvard University, Cambridge, MA 02138, USA}
\email{prineha@seas.harvard.edu}

\makeatother

\begin{document}

\textbf{Single-photon superradiance can emerge when a collection of identical emitters are spatially separated by distances much less than the wavelength of the light they emit, and is characterized by the formation of a superradiant state that spontaneously emits light with a rate that scales linearly with the number of emitters.\cite{Dicke1954,Gross1982,Svidzinsky2008,Roof2016} This collective phenomena has only been demonstrated in a few nanomaterial systems, all requiring temperatures below $10$~K.\cite{Scheibner2007,Tighineanu2016,Raino2018,Grim2019} Here, we rationally design a single colloidal nanomaterial that hosts multiple (nearly) identical emitters that are impervious to the fluctuations which typically inhibit room temperature superradiance in other systems such as molecular aggregates.\cite{Hestand2018} Specifically, by combining molecular dynamics, atomistic electronic structure calculations, and model Hamiltonian methods, we show that the faces of a heterostructure ``nanocuboid'' mimic individual quasi-2D nanoplatelets~\cite{Ithurria2008,Ithurria2011} and can serve as the robust emitters required to realize superradiant phenomena at room temperature. Leveraging layer-by-layer colloidal growth techniques~\cite{Ithurria2012,Hazarika2019} to synthesize a nanocuboid, we demonstrate single-photon superfluorescence \emph{via} single-particle time-resolved photoluminescence measurements at room temperature. This robust observation of both superradiant and subradiant states in single nanocuboids opens the door to ultrafast single-photon emitters and provides an avenue to entangled multi-photon states\cite{Stevenson2006,Dousse2010} \emph{via} superradiant cascades.\cite{Gross1982,Tessier2003}}

Studies of superradiance provide many fundamental insights into the many-body physics of photons, atoms, molecules, and nanomaterials,\cite{Hettich2002,Rohlsberger2010,Roof2016,Solano2017,Hestand2018,Grim2019,Raino2020,Masson2020} and lay foundation for technological applications based on both superradiance and subradiance.\cite{Sipahigil2016,Asenjo-Garcia2017,Rui2020} These applications range from nanomaterial-based lasers~\cite{Jahnke2016} and optical mirrors~\cite{Rui2020} to quantum-optical networks~\cite{Sipahigil2016} and quantum information processing.\cite{Albrecht2019,Wang2020,Head-Marsden2021} This \emph{Article} presents how rational design combined with state-of-the-art synthetic techniques can accelerate the emission rates of colloidal nanomaterials by nearly an order of magnitude \emph{via} superradiance and, arguably more importantly, introduces nanocuboids as a platform to investigate the unique interplay of quantum optics, quantum confinement, and many-body physics at room temperature.

\begin{figure} 
\begin{centering}
\includegraphics[width=15.0cm]{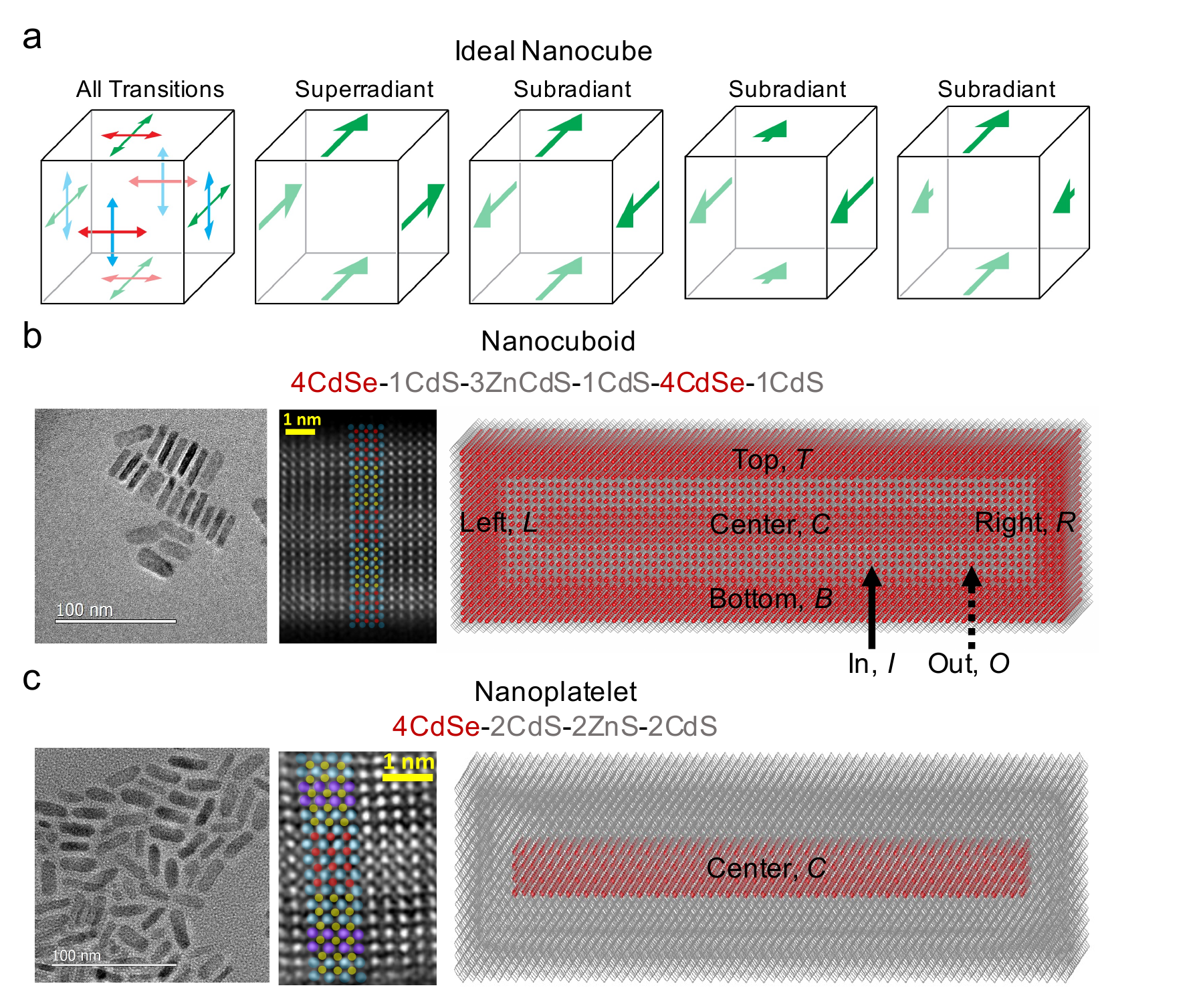}
\par\end{centering}
\caption{(a) The schematic shows a nanocuboid along with arrows displaying the emission polarization of the two lowest energy, optically bright excited states located along each face of the nanocuboid. The colors of the arrows indicate the direction of the emission polarization. The superradiant and subradiant states inherent to a nanocuboid arise from the constructive and destructive interference, respectively, of the transition electric dipole moments between the excited states of the different faces of the nanocuboid, as shown for the $x$-polarized transitions in the schematic. (b) Low-magnification bright field TEM (left) and HAADF-STEM (middle) images are shown alongside a schematic of our nanocuboid in which the active parts (4 monolayers of CdSe, 4CdSe) of the nanocuboid, including the initial 4CdSe seed, are shown in red and the interior and exterior insulating, passivating layers (CdS and ZnS) are shown in gray. The labels are used to identify the $7$ embedded 4CdSe nanoplatelets in the nanocuboid (c) Low-magnification bright field TEM (left) and HAADF-STEM (middle) images are shown alongside a schematic of our single core/shell nanoplatelet in which the active seed (4CdSe) is shown in red and the insulating, passivating layers are shown in gray. The label is used to identify the $1$ embedded 4CdSe nanoplatelet in the core/shell nanoplatelet. The color codes for the HAADF-STEM images in (b) are: Red = Se, Yellow = S, Cyan = Cd/Zn and in (c) are: Red = Se, Yellow = S, Cyan = Cd, Purple = Zn.}
\label{fig:nanocuboid-transitions-structure}
\end{figure}

The observation of superradiant phenomena from a single colloidal nanomaterial is challenging even at low temperatures, requiring multiple identical quantum emitters with large oscillator strengths to be spatially separated by distances on the nanometer scale.\cite{Raino2020,Mattiotti2020} To address these requirements, we specially-designed an epitaxial nano-heterostructure, termed a nanocuboid, to contain multiple quasi-2D nanoplatelets, each with the electronic structure of a quantum well, separated only by a few nanometers. We describe these nanocuboids as a set of coupled nanoplatelets (similar to coupled quantum wells), but they are actually much more complex objects. However, prior to discussing the specifics of our nanocuboids, it is worthwhile to analyze the superradiant and subradiant properties of an idealized nanocuboid, where each face hosts two degenerate excited states localized to its center. We assume that the excited states are optically active with transition electric dipoles of equal magnitude polarized in the plane of each face~\cite{Brumberg2019} (e.g. the $xy$-faces have $x$ and $y$ polarized excitations). This gives a total of twelve excited states (four with $x$-polarizations, four with $y$-polarizations, and four with $z$-polarizations) for a single nanocuboid, as shown in \ref{fig:nanocuboid-transitions-structure}a.

In order to understand the superradiant and subradiant properties of nanocuboids, we invoke a simple and general model Hamiltonian ($H$) to describe its excited states:\cite{Gross1982,Hestand2018}
\begin{eqnarray}
H & = & \sum_{i\alpha}E_{i\alpha}a_{i\alpha}^{\dagger}a_{i\alpha}+\sum_{i\alpha,j\beta}J_{i\alpha,j\beta}\left(a_{i\alpha}^{\dagger}a_{j\beta}+a_{j\beta}^{\dagger}a_{i\alpha}\right).\label{eq:full-model-hamiltonian}
\end{eqnarray}
In equation~(\ref{eq:full-model-hamiltonian}), $i,j,...$ are indices going over all faces of the nanocuboid, $\alpha,\beta,...$ are indices indicating the direction of the transition electric dipole of the excited state, $E_{i\alpha}$ is the energy of the excited state,
$a_{i\alpha}^{\dagger}$ ($a_{i\alpha}$) is a creation (annihilation) operator for the excitonic state $i\alpha$, and $J_{i\alpha,j\beta}$ is the dipole-coupling interaction energy:
\begin{eqnarray}
J_{i\alpha,j\beta} & = & \frac{\left|\vec{\mu}_{i\alpha}\right|\left|\vec{\mu}_{j\beta}\right|}{4\pi\epsilon_{0}\epsilon_{r}\left|\vec{r}_{i}-\vec{r}_{j}\right|^{3}}\left[\hat{\mu}_{i\alpha}\cdot\hat{\mu}_{j\beta}-3\left(\hat{\mu}_{i\alpha}\cdot\hat{r}_{ij}\right)\left(\hat{\mu}_{j\beta}\cdot\hat{r}_{ij}\right)\right].\label{eq:dipole-coupling}
\end{eqnarray}
In equation~(\ref{eq:dipole-coupling}), $\epsilon_{r}$ is the relative permittivity of the host material, $\vec{\mu}_{i\alpha}$ is the transition electric dipole moment, $\vec{r}_{i}$ is the position of the excited state, $\vec{r}_{ij}$ is the vector connecting sites $i$ and $j$, and $\hat{\mu}_{i\alpha}$ and $\hat{r}_{ij}$ are unit vectors. Refer to the Supplementary Information for a discussion of the approximations involved in this model Hamiltonian.

Next we compare the eigenstates predicted by equation~(\ref{eq:full-model-hamiltonian}) for the ideal nanocube described above. Upon diagonalization of equation~(\ref{eq:full-model-hamiltonian}) within the singly excited subspace for this idealized nanocube, three superradiant states are observed, corresponding to the fully symmetric combination of each set of four identical transitions.\cite{Svidzinsky2010} These three superradiant states have radiative decay rates that are $4$ times faster than that of a single emitter (i.e. solving equation~(\ref{eq:full-model-hamiltonian}) with just a single site such as a single core/shell nanoplatelet). Accompanying these three superradiant states are $9$ subradiant states that cannot radiatively decay. \ref{fig:nanocuboid-transitions-structure}a contains visual representations of the one superradiant state and three subradiant states built from linear combinations of the $x$-polarized transitions. Having established the general superradiant capabilities of the surface of a nanocuboid, we will now turn to the specific heterostructure nanocuboid for which we realized single-photon superfluorescence at room temperature.

We designed a nanocuboid by epitaxially embedding multiple quasi-2D CdSe nanoplatelets (NPLs) along the surface of a nanocuboid of insulating CdS and ZnS layers. The nanocuboids shown in \ref{fig:nanocuboid-transitions-structure}b overall dimensions ($L_{x} \times L_{y} \times L_{z}$) of approximately $23\times10\times7$~nm$^3$. Embedded NPLs are a prudent choice for the emitting material for a few reasons. First, NPLs are the fastest fluorescent emitter of any colloidal nanomaterial\cite{Ithurria2011} due to the giant oscillator strength inherent to two-dimensional materials.\cite{Rashba1975} Second, the emission energy and polarization can be precisely tuned by controlling the thickness and lateral dimensions, respectively.\cite{Ithurria2008,Olutas2015} Specifically, the atomically precise control of the NPL thickness results in very strong confinement of a single dimension which causes the emission energy to increase upon increasing the confinement by synthesizing thinner NPLs. NPLs with thicknesses ranging from three to eleven metal layers (MLs) separated by chalcogen atoms can be directly synthesized,\cite{Ithurria2011a,Christodoulou2018} and many more nanostructures can be prepared by the methods of colloidal Atomic Layer Deposition (c-ALD).\cite{Ithurria2012,Hazarika2019} For the case of equal (i.e. square) lateral dimensions, NPLs emit plane-polarized light.\cite{Cunningham2016,Scott2017} However, synthetic protocols can yield rectangular NPLs that can behave similar to quasi-1D nanorods and preferentially emit linearly polarized light.\cite{Cunningham2016,Chen2017} Therefore, NPLs are a unique material because both the energies and polarizations of their excited states can be precisely controlled. This control is vital to the success of realizing superradiant phenomena in multi-NPL colloidal nanomaterials such as the nanocuboids described herein, because superradiance relies on independent and identical emitters interacting with the same photon mode as defined by the energy and polarization of the photon mode.\cite{Gross1982} The final reason is that NPLs embedded within an epitaxial structure are resilient to the fluctuations that typically inhibit room temperature superradiance in molecular systems. In particular, the electron-phonon coupling is more than an order of magnitude weaker in NPLs than in molecular systems, as evidenced by the Stokes shift being on the order of $10$~meV in NPLs and greater than $100$~meV in molecules.\cite{Hestand2018} Additionally, epitaxially embedding the NPLs removes the possibility of misorientation of the transition electric dipoles between neighboring emitters that is known to hinder superradiance in molecular aggregates.\cite{Hestand2018}

For the aforementioned reasons, embedded NPLs are prime candidates for realizing superradiant phenomena at room temperature. The challenge was, thus, to design a synthetic protocol to yield a single nanostructure containing multiple NPLs. We addressed this challenge by utilizing a recently developed layer by layer colloidal growth technique, colloidal Atomic Layer Deposition (c-ALD),\cite{Ithurria2012,Hazarika2019} to synthesize a nanocuboid for which each face of the nanocuboid acts as an independent NPL, mimicking the nanocuboid shown in \ref{fig:nanocuboid-transitions-structure}a. The synthesis of the core/shell NPLs and nanocuboids shown in \ref{fig:nanocuboid-transitions-structure} starts with the synthesis of a colloidal NPL seed (a $4$~ML thick CdSe NPL with lateral dimensions of $17\text{ x }4\text{ nm\ensuremath{^{2}}}$). Next, layer by layer growth was performed via c-ALD to grow passivating shell layers in the case of the core/shell NPL and, in the case of the nanocuboid, passivating layers followed by $4$~MLs of CdSe followed by more passivating layers. The insulating, passivating layers are important for improving the optical efficiencies of colloidal nanomaterials\cite{Hanifi2019} but also introduce strain that can influence the optical properties,\cite{Smith2009,Park2019} as further discussed below. The final core/shell NPLs and nanocuboids are denoted 4CdSe-2CdS-2ZnS-2CdS and 4CdSe-1CdS-3CdZnS-1CdS-4CdSe-1CdS, respectively, where the number prior to the II-VI pair indicates the number of layers grown and the layers are ordered from inner to outer layers (e.g. both nanostructures began with a $4$~ML CdSe core NPL and ended with CdS shell layers). \ref{fig:nanocuboid-transitions-structure} shows the atomistic structures and the structural characterizations of the synthesized core/shell NPLs and nanocuboids. The set of embedded NPLs in the nanocuboids resemble a vertical stack of three quantum wells when considering just the top ($T$), bottom ($B$), and center ($C$) embedded NPLs. However, there is lateral confinement and additional embedded NPLs, labelled in ($I$), out ($O$), left ($L$), and right ($R$) in \ref{fig:nanocuboid-transitions-structure}b, present in our synthesized nanocuboids that are not present in stacks of three quantum wells. The dimensions of each embedded NPL is given in Supplementary Table~S3.

Importantly, the parameters ($E_{i\alpha},~\vec{\mu}_{i\alpha},~\vec{r}_{i}$) in equations~(\ref{eq:full-model-hamiltonian})~and~(\ref{eq:dipole-coupling}) for the specific core/shell NPL ($i\in\{ C \}$) and nanocuboid ($i\in\{ C,T,B,I,O,L,R \}$) shown in \ref{fig:nanocuboid-transitions-structure} can be obtained from atomistic electronic structure calculations, structural characterization techniques, and known scaling laws for NPLs. To this end, we utilized molecular dynamics based structural minimization of the final heterostructure nanomaterials to study the strain and atomistic structure of these materials.\cite{Plimpton1995,Zhou2013} Additionally, a semi-empirical pseudopotential Hamiltonian\cite{Wang1996,Toledo2002} and the Bethe--Salpeter equation\cite{Rohlfing2000} with a static dielectric constant of $\epsilon=5$ were solved to understand the excited states in the core/shell NPL and nanocuboids. These computational techniques have recently been successfully applied to understanding both radiative~\cite{Brumberg2019,Hazarika2019} and nonradiative~\cite{Philbin2018,Philbin2020a} processes of excitons in colloidal nanomaterials. (See the Supplementary Information for more details on the synthetic protocol and computational methods.)

\begin{figure}
\begin{centering}
\includegraphics[width=16cm]{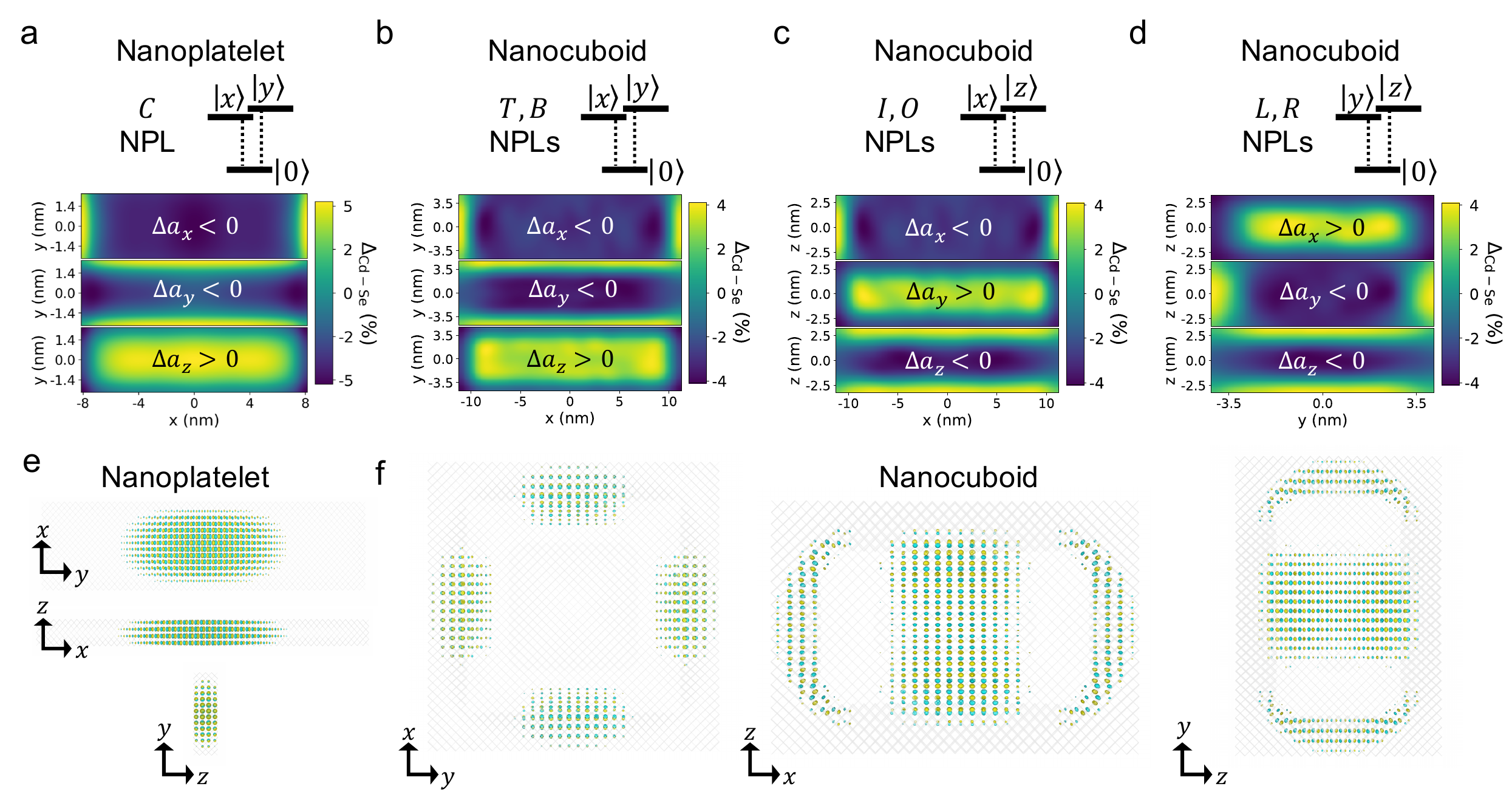}
\par\end{centering}
\caption{Strain profiles in the 4CdSe parts of the nanoplatelet (a, 4CdSe-2CdS-2ZnS-2CdS) and each face of the nanocuboid (b-d, 4CdSe-1CdS-3CdZnS-1CdS-4CdSe-1CdS) are displayed. The set of three plots for each face correspond to the percent change of the CdSe lattice constant ($a$) in the three cardinal directions relative to the ideal projected bond lengths. The percent change of the CdSe lattice constant along the $x$-dimension ($\Delta a_{x}$) is shown on top, $y$-dimension ($\Delta a_{y}$) in the middle, and $z$-dimension ($\Delta a_{z}$) on the bottom. In all cases, the lattice compresses along the large, weakly confined dimensions and expands in the small, confined dimension, demonstrating the biaxial strain in these heterostructure nanomaterials. (e,f) Wavefunctions for the lowest energy hole states are shown for (e) a core/shell nanoplatelet (4CdSe-2CdS-2ZnS-2CdS) with 4CdSe lateral dimensions of $17\times4$~nm\protect\textsuperscript{2} and (f) a nanocuboid (4CdSe-1CdS-3ZnS-1CdS-4CdSe-1CdS) with an initial 4CdSe seed with lateral dimensions $4\times4$~nm\protect\textsuperscript{2}. These wavefunctions are shown to exemplify how the low energy hole states are localized to the 4CdSe parts of the nanocuboids and resemble the states in a single nanoplatelet due to quantum confinement, strain, and band alignment effects.}
\label{fig:atomistic-calculations}
\end{figure}

The low energy excitonic states relevant to emission from the core/shell NPLs (4CdSe-2CdS-2ZnS-2CdS) can be modelled and understood using a simple few-level system, ignoring the exciton fine structure and associated dark states. In Supplementary Table~S4, we explicitly show how almost all of the oscillator strength within the low energy, thermally accessible excitonic states of a single NPL is concentrated within a few ($<10$) excitonic states. Furthermore, there are often only $2$ or $4$ states that contain a majority of this oscillator strength depending on whether the NPL is rectangular or square, respectively. In both cases, all of the transition electric dipole moments of these few low energy bright excited states are orthogonal to the confined axis, just as in each face of the nanocuboid shown in \ref{fig:nanocuboid-transitions-structure}a. The magnitude of the transition electric dipole moments ($\left|\vec{\mu}_{\alpha}\right|$) and the energies ($E_{\alpha}$) of the few important excited states depend on the lateral dimensions of the NPL. The longer lateral dimension will be slightly lower in energy and have a larger transition electric dipole moment (as shown in detail in the Supplementary Information). The 4CdSe NPLs within the core/shell NPL (4CdSe-2CdS-2ZnS-2CdS) have lateral dimensions of $L_{x}=17\text{ nm}$ and $L_{y}=4\text{ nm}$. Thus, the $x$-polarized excited states will be lower in energy and contain more of the oscillator strength than the $y$-polarized excited states (Supplementary Table~S4).  \ref{fig:atomistic-calculations}a contains a schematic of the simplest three-level system that can be used to model the core/shell NPL in this work, and these three-level systems and their extension to few-level ($<10$) systems will be used to model the excited states of the faces of the nanocuboids which will then couple via equation~(\ref{eq:dipole-coupling}) to form superradiant states. 

The connection between the 4CdSe NPL in the core/shell NPL and the faces of the nanocuboid is given credence by noting the similar strain profiles and hole wavefunctions between the embedded 4CdSe NPL in the core/shell NPL and the 4CdSe faces of the nanocuboid shown in \ref{fig:atomistic-calculations}. In particular, biaxial strain, in which the confined dimension expands and the lateral dimensions shrink due to the smaller lattice constants of the CdS and ZnS being grown around the active 4CdSe NPLs,\cite{Hazarika2019} is seen in both the core/shell NPL and each face of the nanocuboid. The mapping of each face of the nanocuboid to NPL is further supported by noting how the low energy excited states are localized to the 4CdSe faces of the nanocuboid. This effect is demonstrated by the hole wavefunctions in \ref{fig:atomistic-calculations}e,f. It is worth noting that in this atomistic electronic structure calculation, the initial center 4CdSe seed in the nanocuboid (4CdSe-1CdS-3ZnS-1CdS-4CdSe-1CdS) has lateral dimensions of only $4\times4$~nm\protect\textsuperscript{2} in order to make the calculation for the entire nanocuboid computationally tractable. The similar strain profiles and band alignment (i.e. confinement) induced excited state localization to the 4CdSe parts of a nanocuboid is key to the mapping of excited states of the nanocuboids to equations~(\ref{eq:full-model-hamiltonian})~and~(\ref{eq:dipole-coupling}) as it allows us to reasonably define the $\vec{r}_{i}$ as the centers of each 4CdSe face of the nanocuboid and the 4CdSe initial center seed.

Another important finding is that within a single face of the nanocuboid, the excitons are delocalized, which is significant for the radiative decay of excitons in NPLs as it, together with the small average electron-hole separations in CdSe NPLs,\cite{Brumberg2019} is responsible for the giant oscillator strengths of NPLs. Mathematically, the radiative decay rate ($k_{\text{rad}}$) of an NPL is proportional to the ratio of the exciton coherence area ($A_{\text{exc}}$) to the square of the exciton Bohr radius ($a_{\text{B,exc}}$, $k_{\text{rad}}\propto A_{\text{exc}} / a_{\text{B,exc}}^{2}$).\cite{Rashba1975,Feldmann1987,Naeem2015} Because the lateral dimensions of all NPLs (i.e. the faces of the nanocuboid) are much smaller than the exciton mean free path in these systems even at room temperature,\cite{Ma2017} the exciton coherence area within an NPL is simply proportional to the area of the NPL ($A_{\text{exc}}\propto A_{\text{NPL}}$). Furthermore, the linear proportionality of the radiative decay rate of an excited state to the lateral area of the face of the nanocuboid is also consistent with a transition electric dipole moment that scales as $k_{\text{rad}}\propto\left|\mu\right|^{2}$.\cite{Feldmann1987} Our electronic structure calculations of 4CdSe NPLs reported in the Supplementary Information also support this scaling.

In terms of equations~(\ref{eq:full-model-hamiltonian})~and~(\ref{eq:dipole-coupling}), the aforementioned results from the molecular dynamics and electronic structure calculations can be summarized in three main points. First, strain and confinement induce exciton localization to 4CdSe of the nanocuboid such that we can reasonably define the positions ($\vec{r}_{i}$) as the center of the six faces of the nanocuboid and the starting center 4CdSe NPL seed to give seven total sites ($i\in\{ C,T,B,I,O,L,R \}$). This demonstrates the connection to the idealized nanocube for which we already detailed its superradiant potential. Second, there are a few ($<10$) low energy excitonic states with large transition electric dipole moments ($\vec{\mu}_{i\alpha}$) in each embedded NPL and the magnitude of the transition dipoles are proportional to the lateral dimensions of the NPL (Supplementary Table~S4).\cite{Rashba1975} And, third, the rigid structure imposed by embedding the 4CdSe NPLs in CdS and ZnS layers results in the geometrical relationship between the transition dipoles of differing faces of the nanocuboid to be constant (i.e. insensitive to thermal fluctuations unlike in neighboring molecules in molecular aggregates). Thus, there is no need to account for fluctuations in the orientation of the transition dipoles when modelling the superradiant properties of the nanocuboids at room temperature. However, disorder in the onsite energies (${\delta}E_{i\alpha}$ where $E_{i\alpha}=E+{\delta}E_{i\alpha}$) can play a central role, as further discussed below.

\begin{figure}
\begin{centering}
\includegraphics[width=15cm]{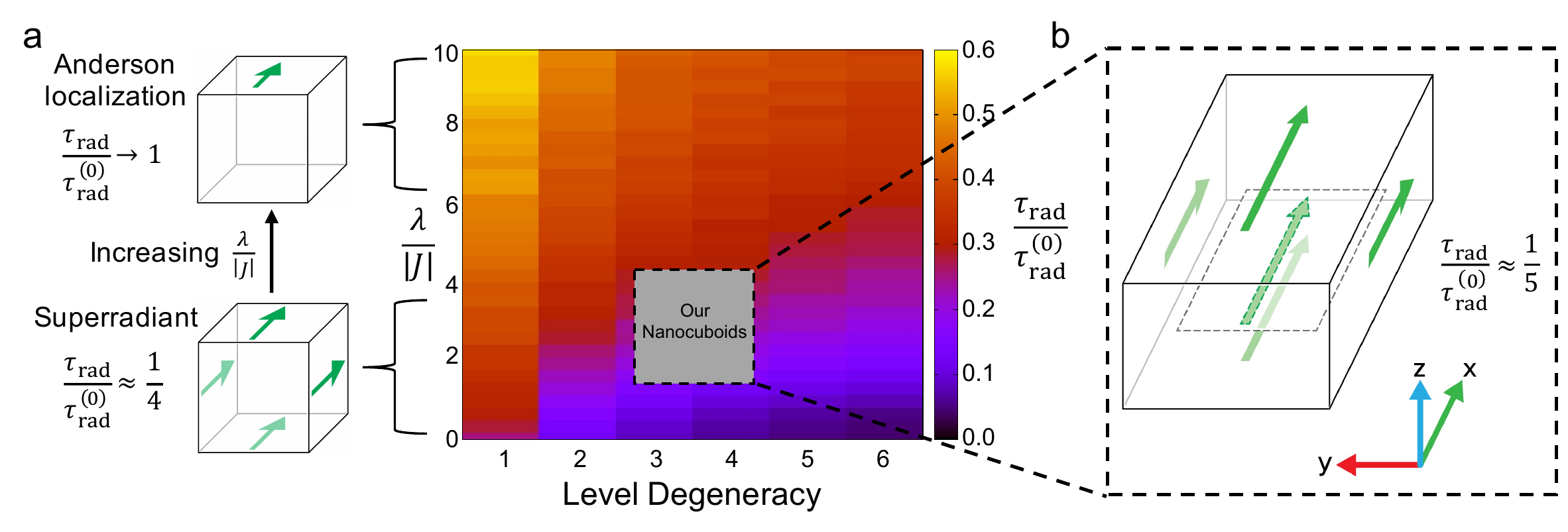}
\par\end{centering}
\caption{(a) The ratio of the fastest radiative lifetime for a nanocuboid (${\tau}_{\text{rad}}$) to that of a single nanoplatelet (${\tau}_{\text{rad}}^{(0)}$) is shown as a function of the level degeneracy and the ratio of the reorganization energy ($\lambda$) to the magnitude of the dipole-dipole coupling strength ($\left|J\right|$). Each pixel represents a different system and $100$ samples were averaged over for each pixel (see the Supplementary Information for more details). The radiative lifetimes of the nanocuboids are shortest for small reorganization energies, defining a ``superradiant regime.'' As the reorganization energy increases, a continuous transition to Anderson-type localization occurs, and the radiative lifetimes of the nanocuboids begin to resemble those of single nanoplatelets. Increasing the number of levels per site (i.e. level degeneracy) makes the superradiant states more resilient to larger energetic fluctuations. Lastly, the location of our synthesized nanocuboids is shown by the gray circle, clearly falling within the ``superradiant regime.'' (b) A visual representation of the superradiant state of our synthesized nanocuboid is shown. The superradiant state is a delocalized excited state in which the $x$-polarized transition electric dipoles of the $C,T,B,I,O$ embedded 4CdSe NPLs constructively interfere.}
\label{fig:phase-diagram}
\end{figure}

To this end, we calculated the radiative decay rates of the excitonic states of nanocubes with various degrees of energetic disorder (${\delta}E_{i\alpha}$) and number of excited state levels per face of the nanocube. The results of these calculations are shown in \ref{fig:phase-diagram}a. The color of each pixel in \ref{fig:phase-diagram}a represents the ratio of the shortest radiative lifetime (averaged over $100$ samples) for a nanocube to that of a NPL with their excited state energies ($E_{i\alpha}$) sampled from a Gaussian distribution with an average energy of $2\text{ eV}$ and a standard deviation equal to the reorganization energy ($\lambda$) and with a fixed dipole-dipole coupling magnitude ($\left|J\right|$) in the nanocubes. The diagram in \ref{fig:phase-diagram}a shows that increasing the disorder in the onsite energies (modelled by increasing $\lambda$) leads to longer radiative lifetimes for a nanocube as a result of Anderson-type localization of the excited states to a single face of the nanocuboid and thus, the radiative lifetimes just resemble those of the individual NPLs (recall that each face of the nanocuboid is similar to an individual NPL). In other words, the delocalized, superradiant states are destroyed by the presence of large energetic disorder (\ref{fig:phase-diagram}a). That being said, increasing the number of excited state levels per face of the nanocuboid stabilizes the superradiant regime, as demonstrated in \ref{fig:phase-diagram}a by the decrease in radiative lifetimes as a function of the number of levels per face. 

Next, we diagonalize the Hamiltonian given by equation~(\ref{eq:full-model-hamiltonian}) using the parameters ($E_{i\alpha},~\vec{\mu}_{i\alpha},~\vec{r}_{i}$) for our specific nanocuboids (4CdSe-1CdS-3CdZnS-1CdS-4CdSe-1CdS) and specific core/shell NPLs (4CdSe-2CdS-2ZnS-2CdS) in order to make predictions on the superradiant properties of these specific structures. Focusing on the single excitation subspace of the Hamiltonian, clear signatures of both superradiant (\ref{fig:phase-diagram}b) and subradiant states are observed (Supplementary Table~S10). Specifically, singly excited states are found that have radiative decay rates that are faster (i.e. superradiant) and slower (i.e. subradiant) in the nanocuboid relative to the core/shell NPL. Quantitatively, there is always a superradiant eigenstate that has a radiative decay that is approximately $5$ times faster than that of the core/shell NPL even when disorder in the onsite energies is taken into account. This signature of single-photon superradiance is highlighted by the gray circle in \ref{fig:phase-diagram}a, where enhancement of the radiative decay rate is primarily due to the superradiant effects that are not destroyed by the onsite disorder and are further promoted by the large number of excitonic states per facet. \ref{fig:phase-diagram}b shows a visual representation of the superradiant eigenstate of our nanocuboid that will be responsible for the short radiative lifetime of our nanocuboid compared to our core/shell NPL. There are also subradiant eigenstates that are either completely or nearly completely dark with radiative decay rates that are $>5$ times slower than that of the core/shell NPL (Supplementary Table~S10). The ideal separation between superradiant and subradiant states discussed above for the idealized nanocube is broken by the disorder (${\delta}E_{i\alpha}$) and the inequivalent positions ($\vec{r}_{i}$) that results in inhomogeneous dipole-dipole coupling ($J_{i\alpha,j\beta}$) between the different 4CdSe NPL faces (Supplementary Table~S3 and Supplementary Fig.~S11). 

\begin{figure}
\begin{centering}
\includegraphics[width=16.5cm]{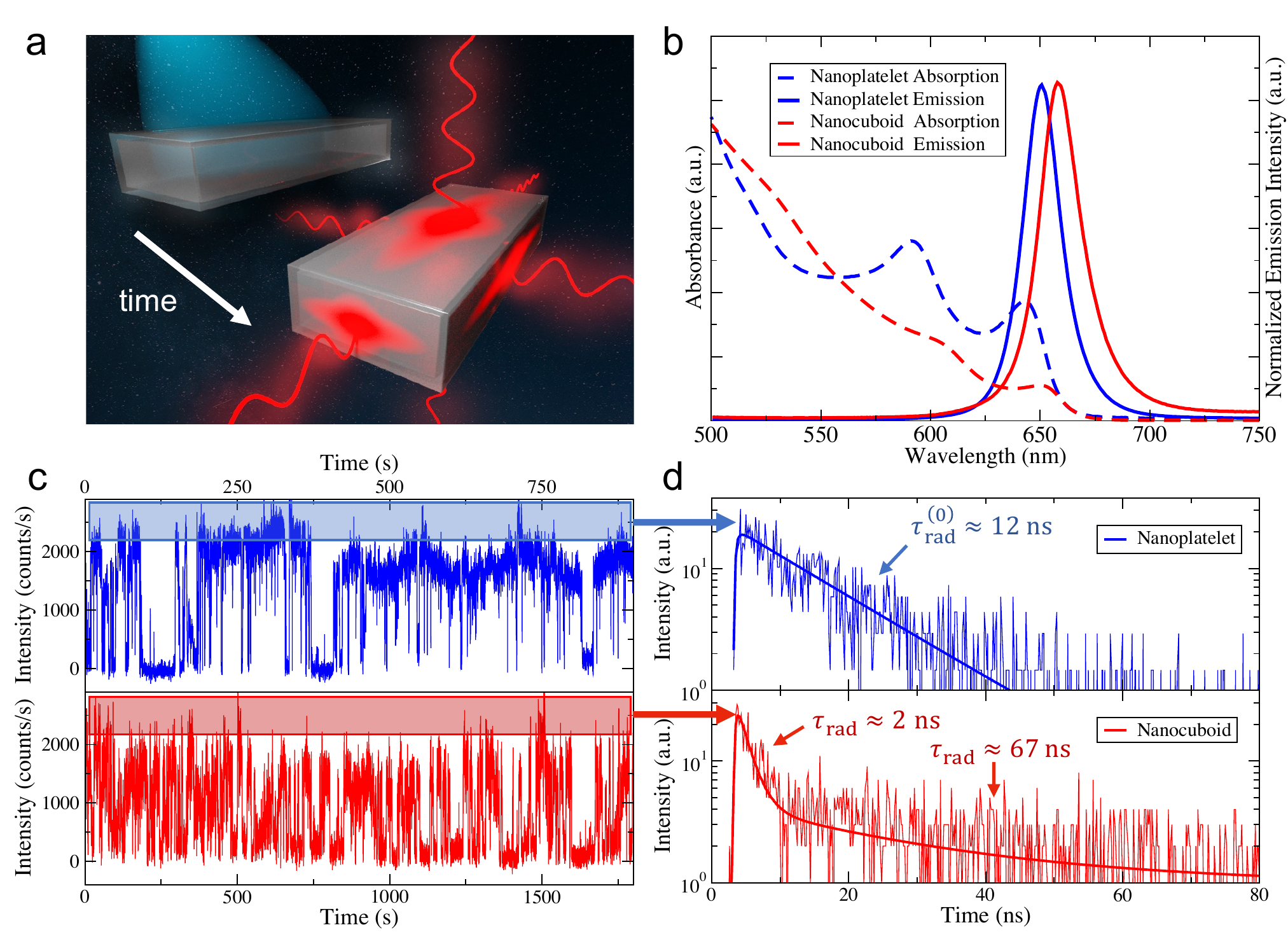}
\par\end{centering}
\caption{(a) A schematic representation of superfluorescence from a nanocuboid. The non-resonant optical excitation that initiates superfluorescence is captured by the blue light shining on a nanocuboid in the top left. The bottom right then shows the symmetric, superradiant excitonic states emitting red light at a future time. The superradiant states are represented by the red ovals along each face of the nanocuboid.  (b) Ensemble absorption and emission spectra for the core/shell nanoplatelets (blue lines) and nanocuboids (red lines). (c) Representative emission time trajectory of a core/shell nanoplatelet (top in blue) and nanocuboid (bottom in red) showing the typical blinking pattern of colloidal nanomaterials. (d) Single-particle time-resolved photoluminescence spectrum extracted from the highly luminescent part of the emission time trajectory. The thick lines are best fits to a single exponential for the nanoplatelet and a bi-exponential for the nanocuboid. The single exponential (thick blue line) has a lifetime of $11.6\text{ ns}$ and the bi-exponential (thick red line) has a fast component with a lifetime of $1.8\text{ ns}$ and a slow component with a lifetime of $67.2\text{ ns}$.}
\label{fig:exp-optical-measurements}
\end{figure}

In order to test the predictions of superradiant and subradiant eigenstates in nanocuboids, we performed time-resolved photoluminescence experiments on the single-particle level as shown schematically in \ref{fig:exp-optical-measurements}a. \ref{fig:exp-optical-measurements} also shows ensemble absorption and emission spectra (\ref{fig:exp-optical-measurements}b) along with representative single-particle blinking (\ref{fig:exp-optical-measurements}c) and time-resolved photoluminescence (\ref{fig:exp-optical-measurements}d) spectra for the core/shell NPLs (4CdSe-2CdS-2ZnS-2CdS) and the nanocuboids (4CdSe-1CdS-3CdZnS-1CdS-4CdSe-1CdS) in blue and red, respectively. We extracted the  reorganization energy ($\lambda$) from the Stokes shift of approximately $20$~meV seen in \ref{fig:exp-optical-measurements}b, which was used as a metric when accounting for the magnitude of the energetic disorder in the final model of our nanocuboids shown in \ref{fig:phase-diagram}. The single-particle blinking spectra show the typical blinking behavior of colloidal nanomaterials.\cite{Krauss1999} Because time-resolved photoluminescence spectra have contributions from both radiative and nonradiative components, we extract radiative lifetimes from just the highly emissive part of the blinking spectrum,\cite{Fisher2004,Brokmann2004,Ma2017} as depicted in \ref{fig:exp-optical-measurements}c,d. 

We find that a single exponential with a radiative decay lifetime of $11.6\text{ ns}$ fits the photoluminescence decay for the core/shell NPL in \ref{fig:exp-optical-measurements}d very well, in agreement with previous room temperature radiative lifetime measurements of core/shell NPLs.\cite{Tessier2013} For the nanocuboid, a bi-exponential function is needed to fit the photoluminescence decay, and there is a clear separation of time scales in the two exponentials. The fast, superradiant component in \ref{fig:exp-optical-measurements}d has a lifetime of $1.8\text{ ns}$ whereas the slow, subradiant component in \ref{fig:exp-optical-measurements}d has a lifetime of $67.2\text{ ns}$. Recall that the diagonalization of equation~(\ref{eq:full-model-hamiltonian}) using the parameters derived from calculations of these exact nanocuboids predicted both superradiant and subradiant excited states with radiative decay rates approximately $5$ times faster (\ref{fig:phase-diagram}b) and $5$ times slower than that of the radiative decay rates of the core/shell NPL. These predictions are in excellent quantitative agreement with the measured lifetimes. Importantly, these radiative lifetimes are reproducible between samples; the Supplementary Information contains the sample to sample variations and demonstrates that the traces presented in \ref{fig:exp-optical-measurements}d are an accurate representation of all the single-particle fluorescent data obtained (Supplementary Fig.~S5 and Fig.~S6). 

An aspect of these measurements that is worth highlighting is that they were performed at room temperature, making nanocuboids one of the first, if not the first, superfluorescent single nanomaterial at room temperature ($298\text{ K}$).  Two more subtle, but very interesting, findings from these experiments is that the observation of both superradiant and subradiant states in the emission dynamics of the nanocuboids suggests that both dynamical dephasing between the faces of the nanocuboid  (which was not accounted for in this work) and phonon-mediated transitions between the superradiant and subradiant states are slow, occurring on the nanosecond timescale. Future theoretical and experimental studies plan to address the mechanism(s) behind these findings.

A brief summary of the properties of the nanocuboids reported herein that were critical to the realization of room temperature superfluorescence from a single nanostructure is warranted. The multi-level nature of the excited states of each emitter (i.e. 4CdSe face) greatly aided in satisfying the energetic resonance conditions between faces required to observe superradiant phenomena. The combination of these many near resonances with the large transition electric dipoles inherent to NPLs and the rigidity of the nanocuboids which allowed us to neglect polarization fluctuations resulted in the delocalized, superradiant excited states in the nanocuboids being stable to the fluctuations inherent to room temperature experiments. 

To conclude, herein we introduced nanocuboids, outlined their superradiant properties, and realized single-photon superfluorescence \emph{via} single-particle time-resolved photoluminescence. Nanocuboids are a promising colloidal nanomaterial due to the state-of-the-art atomic level precision of their synthesis afforded by colloidal atomic layer deposition and the unique interplay of strain and quantum confinement induced exciton localization causing each face of the nanocuboid to act as a nearly identical emitter with large transition electric dipoles of fixed polarization because of the quasi-2D nature of the NPL faces. This localization combined with the non-negligible dipole-dipole coupling between the emitting sites that are spaced only a few nanometers apart results in both superradiant and subradiant states that can be observed \emph{via} single-particle time-resolved photoluminescence experiments at room temperature. 

We envisage future investigations of nanocuboids under higher optical excitation densities might be able to realize entangled photon-pair generation\cite{Stevenson2006,Dousse2010} \emph{via} emission cascades.\cite{Gross1982,Tessier2003} Furthermore, the energy of the emitted photons should be easily tunable by changing the thickness of the emitting NPLs,\cite{Ithurria2011,Ithurria2011a,Christodoulou2018} which is a clear advantage of choosing a quasi-2D material as the emitter. This demonstration of room temperature superfluorescence from a single colloidal nanomaterial is a powerful example of how colloidal synthesis and atomistic modelling can be combined to rationally design new nanomaterials with entirely new quantum optoelectronic functionality, and as a platform to explore previously inaccessible regimes of quantum many-body physics. 

\section{Data Availability}
The data that support the findings of this study are available from the corresponding authors upon reasonable request.

\bibliography{superfluorescence-nanocuboids}

\begin{acknowledgement}
J.P.P., E.R., and P.N. acknowledge helpful feedback from Eli Yablonovitch and Wenjie Dou. 
This work was primarily supported by the Department of Energy, Photonics at Thermodynamic Limits Energy Frontier Research Center, under Grant No. DE-SC0019140. 
This work was partially supported by the Army Research Office MURI (Ab-Initio Solid-State Quantum Materials) Grant No. W911NF-18-1-0431 that supported development of new computational methods. 
The atomistic calculations used resources supported by the Center for Computational Study of Excited State Phenomena in Energy Materials (C2SEPEM), which is funded by the U.S. Department of Energy, Office of Science, Basic Energy Sciences, Materials Sciences and Engineering Division via Contract No. DE-AC02-05CH11231, as part of the Computational Materials Sciences Program and the National Energy Research Scientific Computing Center (NERSC), a U.S. Department of Energy Office of Science User Facility operated under Contract No. DE-AC02-05CH11231. 
The work on nanocuboid synthesis was supported by the Department of Defense (DOD) Air Force Office of Scientific Research under grant number FA9550-18-1-0099 and by NSF under award number CHE-1905290.
This work was performed, in part, at the Center for Nanoscale Materials, a U.S. Department of Energy Office of Science User Facility, and supported by the U.S. Department of Energy, Office of Science, under Contract No. DE-AC02-06CH11357.
X.M. acknowledges support from the Center for Molecular Quantum Transduction (CMQT), an Energy Frontier Research Center funded by the U.S. Department of Energy (DOE), Office of Science, Basic Energy Science (BES).
J.P.P. is a Ziff Fellow at the Harvard University Center for the Environment. 
P.N. is a Moore Inventor Fellow through Grant GBMF8048 from the Gordon and Betty Moore Foundation. 
\end{acknowledgement}

\section{Author Contributions}
The project was jointly conceived by J.P.P., D.V.T., E.R., X.M., and P.N. 
J.P.P. led the theory and calculations, under the guidance of E.R. and P.N. 
J.K. performed the molecular dynamics calculations of the core/shell nanoplatelet and nanocuboid, under the guidance of E.R. 
I.C. and A.H. carried out the experiments on materials synthesis, basic optical and structural characterization and corresponding data analysis, under the guidance of D.V.T. 
L.P. performed the optical measurements of the single nanoparticles and analyzed the experimental data, under the guidance of X.M. 
All authors participated in the data analysis and jointly wrote the manuscript.

\section{Competing Interests}
The Authors declare no Competing Financial or Non-Financial Interests.

\end{document}